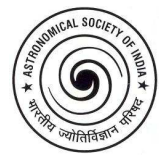

# SoLEXS - A low energy X-ray spectrometer for solar coronal studies

K. Sankarasubramanian*, M. C. Ramadevi, Monoj Bug,
C. N. Umapathy, S. Seetha, P. Sreekumar and Kumar
*Space Astronomy Group, ISRO Satellite Centre, Vimanapura Post, Bangalore 560 017, India*

**Abstract.** A Solar Low-energy X-ray Spectrometer (SoLEXS), a high spectral resolution ($\leq 250$ eV at 5.9 keV) instrument with soft X-ray energy coverage ($\leq 1.5$ keV), is being proposed as an additional payload on-board Aditya-1. The motivation behind this is to complement the visible emission line space solar coronagraph, the main payload on Aditya-1. The science goals in which SoLEXS data will compliment the main payload are: (i) Understanding of DC heating mechanism, (ii) Studies on the Flare-CME relations from the same platform, (iii) Independent and accurate estimates of temperature and emission measure at the flaring sites, and (iv) Coronal abundance studies and its variations during flares. Apart from these four major science goals, this instrument in principle can provide a flare trigger to the main payload and help in optimizing the on-board memory storage.

## 1. Introduction

Aditya-1 is envisaged as a mission to study dynamical nature of the solar corona - the outer layers of the solar atmosphere. Initial studies of these tenuous outer layers were performed only during total solar eclipses in which the high temperature of the corona was first discovered (Edlen 1943) from the green emission line observations. Apart from the high temperature, dynamical phenomenae like Coronal Mass Ejections (CMEs) and flares are often present in the corona. Aditya-1 will also provide data to study CMEs and its initiation mechanism which is yet to be fully understood. Numerous studies on CMEs' properties and its relation to flares were undertaken but conclusive understanding behind their relation is yet to be achieved.

---

*email: sankark@isac.gov.in

We propose a low-energy solar X-ray spectrometer (SoLEXS) for observing flares and hence obtain (i) data for the study of DC heating mechanism, (ii) the physical characteristics of solar flares along with its association to CMEs, (iii) accurate measure of coronal temperature (and Differential emission measure - DEM) as well as abundances of coronal plasma. The proposed payload can also be configured to provide a flare trigger in order to optimize the on-board storage of the main payload.

## 2. Science objectives

Even though solar corona was observed and studied in detail for a long time, the high temperature of the corona is yet to be fully understood. There are two basic mechanisms proposed to explain the high temperature: (1) Waves which can carry energy and dissipate them to heat the corona, known as AC mechanism or wave-heating mechanism and (2) Flares which can happen all over the Sun thought to provide energy for the heating, known as DC mechanism or reconnection theory. While the main payload of Aditya-1 addresses the heating component due to the AC-mechanism, SoLEXS will provide necessary inputs to study the DC component of the heating. It is fairly clear with previous observations that both the mechanism play an important role in the heating processes. It is observed that weaker flare occurs more often than larger flares (Hudson, 1991) and hence weaker flares are the possible candidates for the DC heating mechanism. For sufficient coronal heating the index in the power-law relation between the flare occurrence and its energy has to be larger than -2.0 (Aschwanden 2004; Hudson 1991). The derived power-law index varies from -1.4 to -2.8 obtained with different instruments of varying spectral and spatial resolution (Aschwanden 2004). Although, RHESSI has allowed the study of thermal and non-thermal energy contents in micro-flares systematically, the uncertainty remains in separating the thermal from non-thermal due to instrumental effects and biases (Hannah et al. 2008). To overcome these limitations, an instrument with better energy resolution and lower background is necessary to unambiguously study the energy contents of micro-flares and hence in heating the corona.

There are debates about processes linking flares and CMEs. There is a school of thought that they are different manifestations of the same physical mechanism but observed at different wavelengths. This will be true if the reconnection model proposed is the basic physics behind these two phenomena. Yashiro & Gopalswamy (2008) studied the correlation between different physical quantities relating the X-ray flare observations with the optical observations of CMEs. They find high correlation between the flare fluence and CME kinetic energy emphasizing a common physical mechanism behind flare and CMEs. Their observations suggest that lower the peak x-ray flux of the flare, weaker is the CME association. Their study also confirms that long duration events (LDEs) observed in X-rays usually produce CMEs. However, it may be pointed out that the observations of CMEs carried out in their study is done at or greater than 2 solar radii and hence CMEs with good enough density perturbations at these heights can only be seen. For weaker flares and hence weaker CMEs, the den-sity perturbations may not reach up to 2 solar radii and hence the lower association rate. This observational bias can be removed when CMEs are observed close to the limb with very high sensitive coronagraphs such as the one planned for this mission. Hence, studies of flare-CME relation with Aditya-1 will provide better understanding of the physical mechanisms behind these two dynamical phenomenae.

Table 1. Table of Requirements for SoLEXS.

| Properties | Desired | SoLEXS' Capabilities |
|---|---|---|
| Energy Range (>5% efficiency) (keV) | 0.5 - 30 | 1.5 - 30 |
| Spectral Resolution (at 5 keV) | ≤ 250eV | 250eV |
| Angular Resolution | few arcsec | Non-imaging |
| Dynamic Range | << A1 to X100 flares | < A1 to X-class flares |
| Max Counting Rate | 500,000 | 500,000 |
| Radiation Hardness | sustain $10^9$ protons with only 10% loss in resolution | |
| Area of the Aperture (for C to > X-class) | | 0.1 sq-mm |
| Area of the Aperture (for < A to B-class) | | 5 sq-mm |
| Detector Temperature | -30 deg. C | -20 deg. C |
| Calibration Source | > two energies | two energies |

## 3. SoLEXS requirements and configuration

The flux of the Sun in the low energy range of around few keV, is very high ( $10^6$ photons mm$^{-2}$ s$^{-1}$) and variable. The solar X-ray spectrum is generally soft, with most of the photons concentrated at energies around and below 3 keV. The variability, during flares, is mostly at higher energies associated with variable high temperature. Photon flux for an A-class flare is typically 20-photons/sec/sq-mm where as for X-class flare it is $2 \times 10^5$ photons/sec/sq-mm over 1 to 10 keV and hence requires a detector with large dynamic range capability. With the current technology, a single detector cannot provide such a large dynamic range. SoLEXS is configured with a single detector along with a variable aperture mechanism to cover the required dynamic range. To study the detailed emission mechanisms and abundances during flares, a high spectral resolution is a must. Most of the spectral lines present in the solar corona are at the soft X-ray regime, below 10 keV. More crowded lines are seen below 3 keV, hence spectral resolution of atleast 250eV will be required in this regime. SoLEXS is aimed at providing a spectral resolution ≤ 250 eV@5.9 keV and observe low energies down to <1.5 keV in order to observe as many spectral lines as possible.

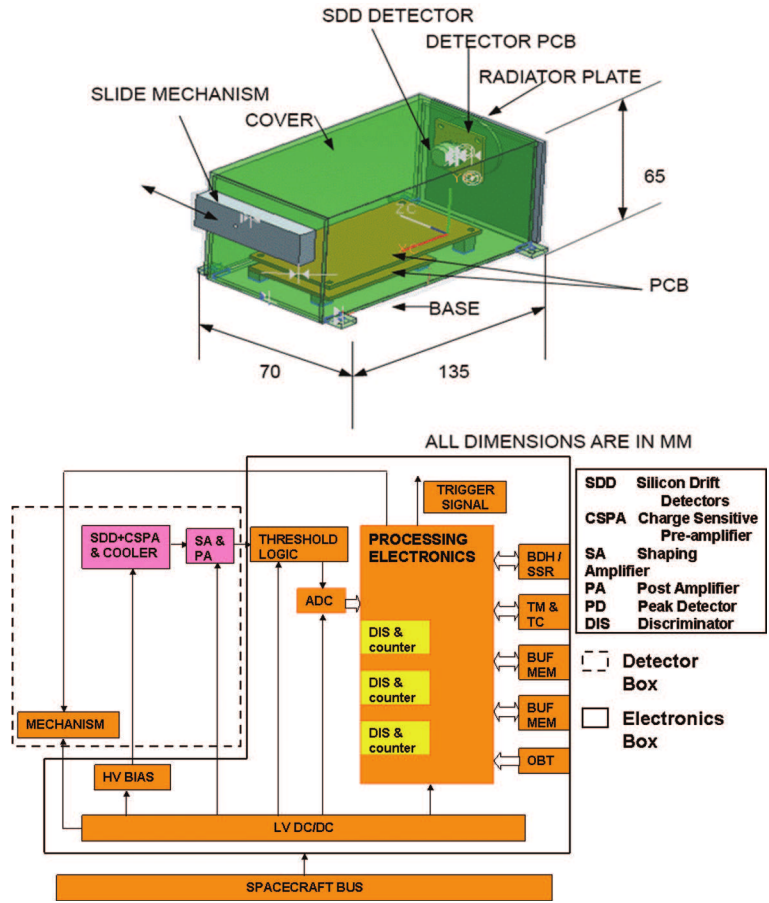

**Figure 1.** Top: Schematic block diagram of the detector box. Bottom: Block diagram of the processing electronics.

Silicon Drift Detector (SDD) is the proposed detector for SoLEXS. SDDs are new type of detectors which are Si-based with low anode-capacitance and low leakage current, yielding good spectroscopic performance. They are chosen mainly because of its large dynamic range capability as well as spectral resolution under nominal cooling compared to the conventional Si-PIN used in earlier missions. SDDs are tested for performance in laboratories and flown once on-board NASA's APXS in Mars Exploration Rover. The maximum possible count rate without showing any degradation in the spectroscopic performance is about 500,000 cps for SDDs an order more than Si-PIN. The efficiency at lower energies can be low due to the entrance window used to cut-off visible light. For a 12.5 micron Be window, the efficiency is about ≈20% at around 1 keV. The efficiencey at 30 keV will fall down to ≈30% for a 450$\mu$m active depth of the detector. Detector cooling is acheived using Peltier cooler integrated within the detector. Table 1 lists out the instrument parameters of SoLEXS. A mechanism is used to select the required aperture. SoLEXS will consist of two boxes: detector box and electronics box. Fig. 1 (left) shows the schematic diagram of the detector box along with the pre-amplifier unit and the bias board to power the detector. It would also include a mechanism for the aperture change. The mechanism will have three selectable position, (i) small aperture (≈0.1mm$^2$), (ii) large aperture (≈5mm$^2$), and (iii) opaque with a calibration source. This box will be mounted on the sun facing side of the satellite panel with the detector pointing towards the Sun. Since this box will always face the Sun, thermal modeling is important in order to keep the detector box to within 30$^o$C so that the detector achieves -20$^o$C through its peltier cooler. The electronic box includes the front-end electronics, the processing electronics and the DC-DC converters required to power up the detector as well as other electronics. Figure 1 (right) shows the schematic diagram of the electronics chain. This chain of electronics will provide the light curve of the X-ray flux as well as the spectra. The counter used will provide us with the count rates which is then used to provide the flare trigger as well as the mechanism movement. The details of the resource requirements for SoLEXS is provided in Table 2.

Table 2. Resource requirements.

| Parameter | Total |
|---|---|
| Mass (in kg) | < 2 + Mechanism weight |
| Size (L X W X H) (in mm) | Box1: 150 × 60 × 60 |
| | Box2: 180 × 120 × 70 |
| | Boxes can be mounted separately (separation ≤ 1m) |
| Regulated Power (in W) | 12.25 + Mechanism power |
| Raw power (in W) | 17.5 + Mechanism power |
| | about 2W can be saved if 1553 bus can be shared with the main payload |
| Storage | 350 Mbits in 12-hour period |

## 4. Laboratory developments

A laboratory model of this payload was developed and tested at the Space Astronomy Group. SDDs along with the pre-amplifiers and integrated Peltier cooler were procured. Thermal control of the SDD was developed in-house to operate the detector at temperatures below -10$^o$C when the ambient was at room temperature. The analog and digital processing cards were developed in-house to obtain the spectral and timing information of the incident X-ray photons. Laboratory testing was carried out to demonstrate the achieved spectral resolution and low-energy threshold. Figure 2 shows the preliminary laboratory testing of the SDD detector module. Laboratory power supplies were used for this experiment. The K-alpha and K-beta lines of the calibration source Mn, were clearly separated in the plot. The achieved spectral resolution is $\approx$200eV at 5.9keV and the achieved low-energy threshold was 1.1 keV. Efforts are underway to replace the laboratory power sources with DC-DC converters to obtain an end-to-end testing of the laboratory module.

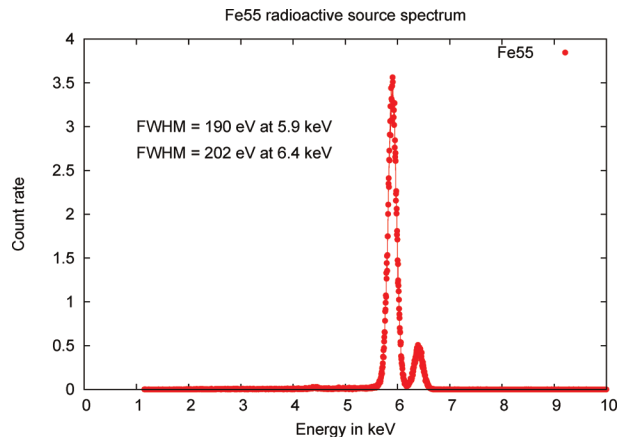

**Figure 2.** Top: Laboratory test setup of the SDD detector. Bottom: Observed Fe55 source spectra showing the Mn K-$\alpha$ and K-$\beta$ spectra clearly separated.

## 5. Summary

A solar low-energy X-ray spectrometer is being proposed to provide complimentary data to the visible emission line space solar coronagraph - main payload of Aditya-1 mission. This instrument will observe all classes of solar flares and hence provide required data for the DC heating mechanism. Being on the same satellite, these two instrument can provide high timing accuracy data for the study of CME-flare relations. The accurate flare temperature estimates due to the high spectral resolution data will be useful for the study of flare dynamics. Apart from providing high spectral resolution data for coronal studies, SoLEXS can also provide a flare trigger to the main payload for operation in CME mode. The coronagraph instrument is expected to operate in CME mode for > 50% of its lifetime. In this mode, it produces a large volume of data ($\approx$21 Gb per day) and hence fill up the on-board memory quickly. An on-board decision of discarding non-CME data can be established with SoLEXS since all CMEs have associated flares with varying strength.